\documentclass[10pt,twocolumn,letterpaper]{article}

\usepackage[pagenumbers]{iccv} 

\definecolor{iccvblue}{rgb}{0.21,0.49,0.74}

\usepackage[pagebackref,breaklinks,colorlinks,allcolors=iccvblue]{hyperref}
\usepackage{tikz}
\usepackage[most]{tcolorbox}
\usepackage{svg}
\usepackage{soul}
\usepackage{wrapfig}
\usepackage{bm}
\usepackage{adjustbox}
\usepackage{pifont}
\usepackage{graphicx}
\usepackage{amssymb}
\usepackage{booktabs}
\usepackage{makecell} 
\usepackage{multirow}
\usepackage{multicol}
\definecolor{Gray}{gray}{0.9}
\definecolor{myblue}{RGB}{230, 230, 248}
\definecolor{featherbrown}{HTML}{8C564B}
\definecolor{featherbackgroundbrown}{HTML}{F1E1DB}
\definecolor{fastvgreen}{HTML}{3B7D23}
\definecolor{fastvbackgroundgreen}{HTML}{EDF8E7}
\definecolor{localizationpurple}{HTML}{78206E}
\definecolor{localizationbackgroundpurple}{HTML}{F8E7F6}
\usepackage{amsmath}
\usepackage{enumitem}
\usepackage{algorithm}
\usepackage{algpseudocode}

\usepackage{colortbl}
\hypersetup{
	colorlinks=true,
	citecolor=teal,
}

\def\paperID{*} 
\def\confName{ICCV}
\def\confYear{2025}

\newcommand{\ModelName}{S2C-Diffusion}
\title{Separate to Collaborate: Dual-Stream Diffusion Model for Coordinated Piano Hand Motion Synthesis}

\author{
Zihao Liu\footnotemark[1],~ 
Mingwen Ou\footnotemark[1],~
Zunnan Xu\footnotemark[1],~
Jiaqi Huang,~
Haonan Han,~
Ronghui Li,~
Xiu Li\footnotemark[2]
\\
Tsinghua University
}

\begin{document}
\maketitle

\footnotetext[1]{Equal Contribution.} 
\footnotetext[2]{Corresponding author.}

\begin{abstract}
Automating the synthesis of coordinated bimanual piano performances poses significant challenges, particularly in capturing the intricate choreography between the hands while preserving their distinct kinematic signatures. In this paper, we propose a dual-stream neural framework designed to generate synchronized hand gestures for piano playing from audio input, addressing the critical challenge of modeling both hand independence and coordination. Our framework introduces two key innovations: (i) a decoupled diffusion-based generation framework that independently models each hand's motion via dual-noise initialization, sampling distinct latent noise for each while leveraging a shared positional condition, and (ii) a Hand-Coordinated Asymmetric Attention (HCAA) mechanism suppresses symmetric (common-mode) noise to highlight asymmetric hand-specific features, while adaptively enhancing inter-hand coordination during denoising. Comprehensive evaluations demonstrate that our framework outperforms existing state-of-the-art methods across multiple metrics. Our
project is available at \href{https://monkek123king.github.io/S2C_page/}{S2C}.
\end{abstract}
    
\begin{figure}
  \includegraphics[width=0.45\textwidth]{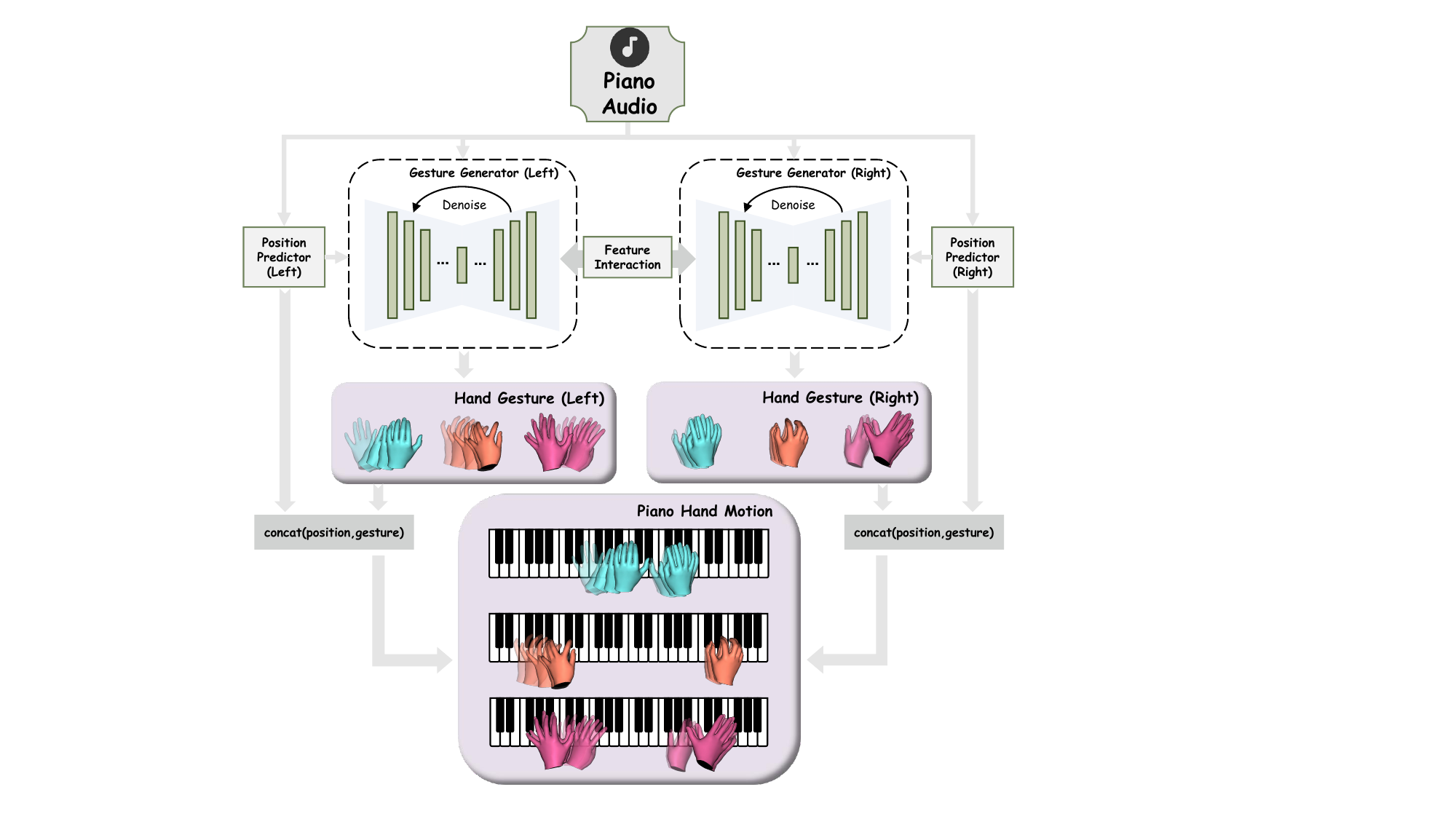}
  \caption{Given a segment of piano audio, our proposed method first decouples the motion generation of the left and right hands and then employs a coordination mechanism to model their interactions. This “separate to collaborate” strategy enables the synthesis of natural, expressive, and well-coordinated piano hand motions.}
  \label{fig:show}
\end{figure}

\section{Introduction}
Driven by advances in artificial intelligence, generating natural and precise human motion from multimodal inputs has become a prominent and promising field of research. Among various input modalities, audio-driven motion synthesis~\cite{Alexanderson_2023,yang2024freetalker,xu2024mambatalk,hong2025audio} has gained increasing attention, particularly in applications related to embodied intelligence and virtual reality~\cite{li2024dispose,jin2024alignment,ma2025msdetreffectivevideomoment}. As an emerging but impactful technology, it has been widely adopted in areas such as gaming, animation, and webcast, fueling innovation and improving user experiences across these domains~\cite{han2024reparo,lin2024consistent123,xu2024chain,li2024lodge,lin2025mvportrait,ma2025finegrainedzeroshotobjectdetection,xu2025hunyuanportrait}. A rising subfield within this area is piano-hand motion generation~\cite{zakka2023robopianist,wang2024furelisecapturingphysicallysynthesizing,gan2025pianomotion10mdatasetbenchmarkhand}, which has garnered growing research interest in recent years. This technology holds promise for developing AI-powered piano tutors that can provide learners with valuable guidance on fingering and rhythm techniques. However, piano performance requires highly complex and coordinated hand movements that involve precise alignment with musical timing, intricate finger trajectories, spatial hand positioning, and fine inter-hand coordination. Traditional rule-based methods~\cite{kranstedt2002murml, Ravenet2018AutomatingTP} struggle to capture the natural fluidity of such motions, while existing data-driven approaches~\cite{peng2023emotalk,zhi2023livelyspeaker,gan2025pianomotion10mdatasetbenchmarkhand}, though more expressive, often suffer from such problems as temporal discontinuity and lack of fine-grained control. Consequently, it is of both theoretical and practical importance to explore methods that can accurately model the relationship between audio and hand motion and synthesize realistic, coordinated gestures for piano performance.

In recent years, the field of fine-grained hand motion generation has made remarkable progress. Numerous works have explored dexterous hand control and physically plausible hand-object interactions. For example, Chen et al.~\cite{Chen_2023} propose a method to synthesize physically realistic nonprehensile pregrasp motions. Their approach automatically discovers diverse strategies for leveraging environmental contacts to assist hand movements. Andrychowicz et al.~\cite{andrychowicz2020learning} employ reinforcement learning (RL) to train vision-based object reorientation policies on a physical Shadow Dexterous Hand, enabling precise in-hand manipulation. Yang et al.~\cite{yang2022learning} focus on chopstick-based object relocation tasks, which require intricate and coordinated hand control. Zhang et al.~\cite{zhang2021manipnet} propose a hand-object spatial representation that enables generalization from limited data. Mordatch et al.~\cite{mordatch2012contact} present a method for the automatic synthesis of dexterous hand movements based on high-level task goals, such as grasping and picking up objects. While these methods have advanced the understanding and generation of complex hand behaviors, they typically focus on single-hand scenarios and often overlook bimanual coordination. However, many real-world tasks, such as playing the piano, require the fine-tuned collaboration of both hands. Piano performance is a representative example of a domain where both independence and coordination between the left and right hands are crucial. During piano playing, each hand follows a distinct yet complementary motion pattern, which makes the generation of realistic bimanual gestures especially challenging. One of the most representative works in this domain is PianoMotion10M~\cite{gan2025pianomotion10mdatasetbenchmarkhand}, which introduces a large-scale paired dataset of piano audio and 3D hand motion, along with a benchmark model and a set of evaluation metrics for assessing motion quality. Nevertheless, their method adopts a single-stream architecture to model both hands jointly, which fails to account for the asymmetric roles and independent movement dynamics of the left and right hands. As a result, the generated motions often lack naturalness and inter-hand coordination, limiting their realism and applicability in real-world virtual performance systems.

To address the limitations of existing methods, particularly the inability to effectively capture the independence and coordination of both hands during piano performance, we propose a diffusion-based framework for piano hand motion synthesis. 
Our approach centers around a dual-stream architecture that explicitly separates the modeling processes for the left and right hands.
Specifically, we employ two independent diffusion models to learn the motion patterns of each hand separately. This design enables the system to better preserve the distinct kinematic features and temporal dynamics unique to each hand, which are often overshadowed in single-stream architectures. 
Our approach introduces a novel Hand-Coordinated Asymmetric Attention (HCAA) mechanism to improve coordination between the two hands. This mechanism uses an asymmetric strategy to reduce noise interference during the denoising process of diffusion models. More importantly, it creates dynamic interactions between the two generative streams, allowing them to share contextual cues and stay aware of each other over time. This ensures that the generated bimanual motions are both temporally consistent and spatially aligned, resulting in smoother and more natural interactions between the two hands.
In addition, we incorporate a decoupled position prediction module that supplies accurate 3D spatial information for both hands. This module utilizes a position-sharing mechanism to guide both diffusion streams with a shared understanding of global hand locations, which further strengthens their interdependence and temporal consistency. 
The motivation behind our design is grounded in the nature of piano performance: the two hands operate both independently, each responsible for different melodic or harmonic roles, and cooperatively requiring tight temporal and spatial synchronization. 
Traditional unified models often have difficulty balancing these dual characteristics. Our dual-stream diffusion framework, combined with the HCAA module, focuses on capturing the complex interaction between the hands. This leads to more natural, expressive, and coordinated hand movements for piano playing.
Our paper presents a novel framework for generating natural and synchronized piano hand motions directly from audio inputs, as illustrated in Fig.~\ref{fig:show}. We use a dual-stream diffusion model that separately creates motion for each hand. This model is combined with a Hand-Coordinated Asymmetric Attention mechanism, which improves denoising and hand coordination. Our pipeline also includes position-based motion synthesis, where audio features and predicted hand positions together guide the motion creation process.
The main contributions can be summarized as follows: 
\begin{itemize}[leftmargin=*,noitemsep,nolistsep]
    \item We propose a decoupled diffusion-based architecture that separately models the motion of the left and right hands, capturing their distinct dynamics.
    \item We design a Hand-Coordinated Asymmetric Attention (HCAA) mechanism to improve motion coherence and coordination during the denoising process.
    \item Extensive experiments on the PianoMotion10M dataset demonstrate that our method outperforms state-of-the-art baselines in terms of motion naturalness, temporal consistency, and hand synchronization.
\end{itemize}
\section{Related Work}
\subsection{3D Human Motion Synthesis}
3D human motion synthesis aims to generate natural human pose sequences and has significant potential for practical applications. Recent advancements in the field of human motion generation can be divided into two subcategories: autoregressive-based and diffusion-based methods.
Autoregressive-based methods~\cite{guo2022tm2t,lu2023humantomatotextalignedwholebodymotion,guo2023momaskgenerativemaskedmodeling,yang2023qpgesture,zhang2023t2mgptgeneratinghumanmotion,chen2025focused,zhong2023attt2mtextdrivenhumanmotion,han2024atom,lyu2025hvis} capture the temporal dependencies in action sequences through frame-by-frame prediction, thereby enabling the generating of complex action sequences. 
For example, MotionGPT~\cite{jiang2023motiongpthumanmotionforeign} treats human motion as a language and generates coherent motion sequences using large-scale pre-trained models. However, this kind of method is susceptible to error accumulation during the generation of long sequences, resulting in certain deficiencies in the overall quality and stability of the generated actions.

To address this problem, diffusion-based model~\cite{chen2023executingcommandsmotiondiffusion,luo2023latentconsistencymodelssynthesizing,li2023finedance,jin2023actwishfinegrainedcontrol,li2024interdance,kong2023prioritycentrichumanmotiongeneration,li2024exploring,wang2023fgt2mfinegrainedtextdrivenhuman,li2024lodge++,mi2025data,lu2025adversarial,zhang2022motiondiffusetextdrivenhumanmotion,lee2024interhandgen} generation method was introduced into human motion generation. The diffusion model~\cite{ho2020denoisingdiffusionprobabilisticmodels,song2022denoisingdiffusionimplicitmodels,yang2023diffusestylegesture,yang2023unifiedgesture} generates samples by reversing the noise process of the long Markov chain, which can effectively alleviate the error accumulation problem while maintaining motion coherence. Representative works include the Motion Diffusion Model (MDM)~\cite{tevet2022humanmotiondiffusionmodel}, which uses a classifier-free guidance mechanism to achieve high-quality motion generation; And Guided Motion Diffusion (GMD)~\cite{karunratanakul2023guidedmotiondiffusioncontrollable}, by combining text prompts, reference trajectories, and key positions,  achieves controllability and diversity of the generation process. The diffusion model has shown advantages in generating high-quality and continuous motion sequences, and has become an important research direction in the current field of human motion generation. 
It should be noted that in the subfield of piano-audio-driven hand motion generation, recent studies such as pianomotion10M~\cite{gan2025pianomotion10mdatasetbenchmarkhand} have attempted to apply diffusion models to hand motion synthesis in piano playing scenarios. However, existing methods generally model bimanual movements as a joint representation. This kind of strong coupling assumption ignores the functional differentiation and asynchrony of the left and right hands during performance in the real world. To address these key issues, we propose a dual-stream decoupled generation framework, which improves the detailed expression and musical fit of the generated hand motions while maintaining the coordination of both hands' movements by designing independent conditional diffusion models and feature interactive attention mechanisms.

\subsection{Piano Guided Hand Motion Generation}
The study of physically simulated dexterous hand control strategies has many applications in computer graphics and embodied intelligence. Most of the existing research~\cite{Chen_2023,liu2008synthesis,mordatch2012contact,wang2013video,yuan2023physdiff,andrychowicz2020learning,xie2023hierarchical,yang2022learning,zhang2021manipnet} on physically based dexterous hand control focuses on one-handed tasks and requires relatively low control accuracy. Our study focuses on the generation of piano-playing motion, a challenging task that requires synergistic control of both hands and a high level of temporal and spatial accuracy.

Playing the piano is a common but complex physical activity in daily life, and the associated movement generation guided by piano audio has gradually attracted much attention. Traditional strategies for learning to play the piano in simulation through reinforcement learning~\cite{xu2022towards,zakka2023robopianist}. With the breakthrough of the diffusion model in the field of motion generation, recent studies~\cite{wang2024furelisecapturingphysicallysynthesizing,gan2025pianomotion10mdatasetbenchmarkhand} have begun to explore its application to the piano-audio-driven hand motion generation. For example, FürElise~\cite{wang2024furelisecapturingphysicallysynthesizing} showed good results on a dataset of 10 hours of 3D hand motion with audio data through the diffusion generation and motion retrieval framework. Especially, the PianoMotion10M dataset~\cite{gan2025pianomotion10mdatasetbenchmarkhand} provides a new benchmark for hand motion generation in piano performance, with 116 hours of 3D hand motion with audio data and 10 million hand gesture annotations. However, existing methods model both hands as a whole during the generation process, failing to take into account the significant differences between the left and right hands in terms of task division, movement patterns, and rhythms, leading to distortions in the generation process.
Therefore, we propose transitioning from general hand modeling to fine-grained hand decoupling modeling for piano motion generation. In this decoupling generation process, ensuring mutual perception between the left and right hands becomes another key issue. Therefore, we introduce an asymmetric feature interaction mechanism that constructs a left-right asymmetric feature space to explicitly suppress common-mode (symmetric) noise~\cite{laplante2018comprehensive}, thereby amplifying differential-mode (asymmetric) information and facilitating the dual-stream decoupled generation architecture in balancing motion independence and coordination.

\begin{figure*}[t] 
  \includegraphics[width=\textwidth]{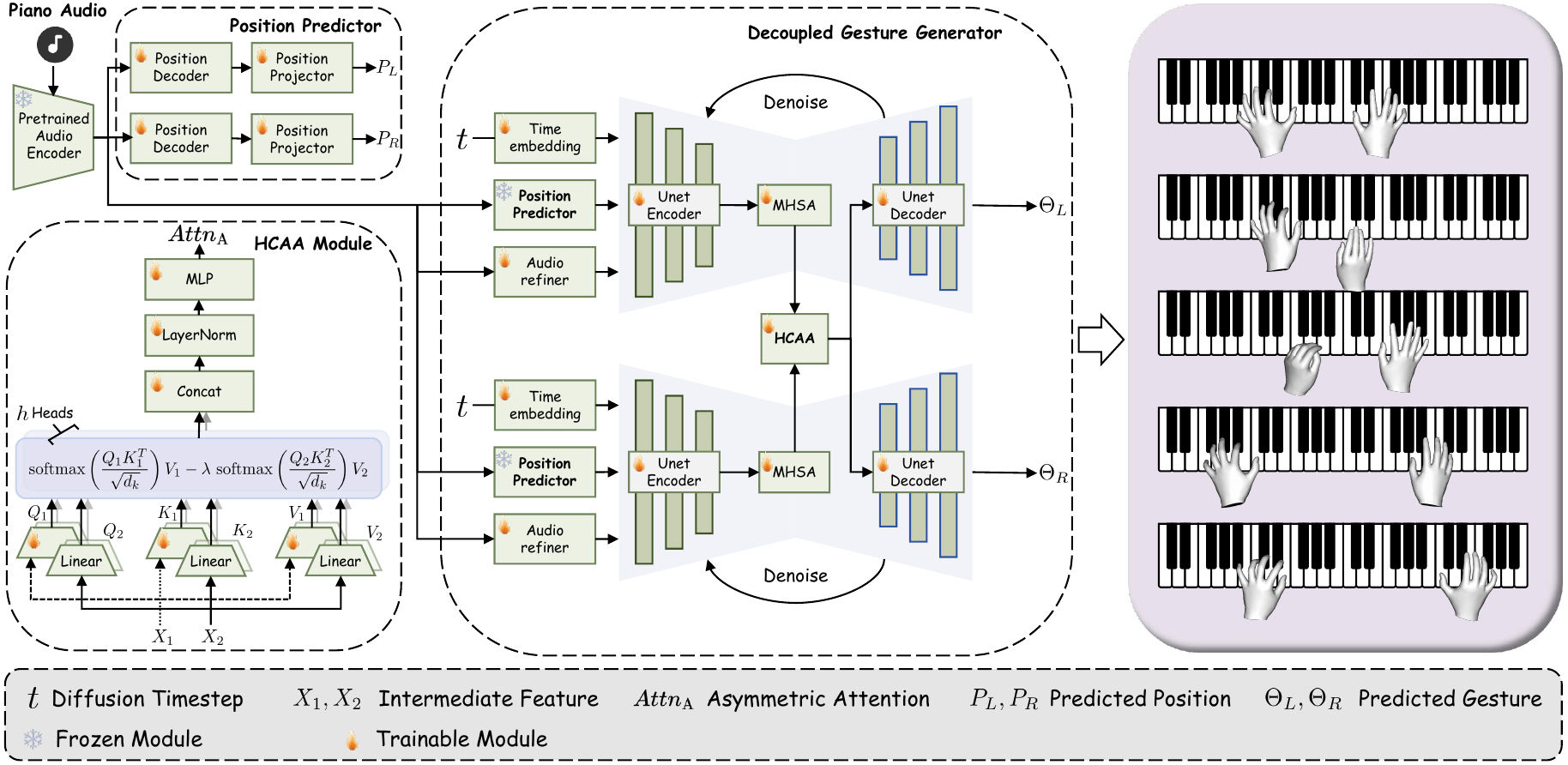}
  \caption{The overall framework of our method. It consists of three main modules: (1) Position Predictor, which predicts hand positions from piano audio; (2) Hand-Coordinated Asymmetric Attention (HCAA), applied to the intermediate U-Net features after a Multi-Head Self-Attention (MHSA) layer, enabling asymmetric feature interaction between the two hands. (3) Decoupled Gesture Generator, which employs two diffusion models to generate motions for the left and right hands.}
  \label{fig:framework}
\end{figure*}
\section{Method}
In this section, we introduce our dual-stream diffusion model. Our framework, as shown in Fig.~\ref{fig:framework}, is composed of three key components: a position predictor that estimates hand positions from piano audio, a feature interaction module built around Hand-Coordinated Asymmetric Attention, and a decoupled piano hand gesture generation module using diffusion models. We begin by defining the problem and outlining the training process, then proceed to gesture generation, and finally illustrate the feature interaction module.

\subsection{Problem Formulation}
Given a N frames piano audio piece $A = \{a_1,...,a_N\}$, our goal is to generate the corresponding position sequence $P = \{p_1, ..., p_N\}$ and hand gesture sequence $\Theta = \{\theta_1, ..., \theta_N\}$, where $ p_i\in \mathbb{R}^{3\times2}$ and $\theta_i \in \mathbb{R}^{J \times 3 \times 2}$, with $J$ denoting the total number of hand joints (48 per hand in our implementation). The hand position sequence P represents the 3D coordinates of both the left and right hands at each frame, while the hand gesture sequence $\Theta$ captures the Euler angles of each joint, providing a detailed representation of the dynamic hand movements.

\subsection{Training Pipeline}
Our proposed method for generating piano hand motions is based on a diffusion model and features a decoupled architecture. The training process involves three main components: audio encoding, training of position predictors, and training of gesture generators.

\textbf{Audio Encoding.} Firstly, the input piano audio $A$ is processed through a pretrained audio feature extractor $\Phi_a$ to obtain a compact representation of the audio features $f_a$ . This process is defined as $f_a = \Phi_a(A)$, where $f_a \in \mathbb{R}^{N \times C}$ and $C$ denotes the feature dimension of $\Phi_a$ . This representation encapsulates essential auditory features, providing high-quality input for subsequent motion generation.

\textbf{Stage 1: Training the Position Predictor.} The objective of the first training stage is to train a position predictor $\Phi_{A2P}$, which predicts hand position from the extracted audio features $f_a$. Specifically, to learn the mapping between audio and hand positions for both hands accurately, we train separate position predictors for the left and the right hand. The outputs of these models are given by: $P_L = \Phi_{A2P}^L(f_a)$, $P_R = \Phi_{A2P}^R(f_a)$, where $P_L, P_R \in \mathbb{R}^{N \times 3}$. After completing the first training stage, the weights of the position predictors are frozen to ensure that the generated positional information remains stable in the subsequent stages.

\textbf{Stage 2: Training the Motion Generator.} This stage focuses on training the gesture generator $\Phi_{A2M}$ to synthesize complete hand gesture sequences conditioned on audio features $f_a$ and predicted positional information $(P_L, P_R)$. 
During the inference phase, the diffusion-based generator takes these multimodal inputs to produce natural piano performance motions through iterative denoising. Similar to the position predictor's architecture, we implement a decoupled bilateral training strategy: $\Theta_L = \Phi_{A2M}^L(f_a, P_L, P_R)$, $\Theta_R = \Phi_{A2M}^R(f_a, P_R, P_L)$, where $\Theta_L, \Theta_R \in \mathbb{R}^{N \times J \times 3}$, and J represents the number of hand joints. 
Additionally, our method integrates a feature interaction module into the diffusion model's denoising network. This architecture allows us to separately model hand-specific kinematics while maintaining inter-hand coordination through cross-parameter conditioning.

\subsection{Decoupled Hand Motion Generation}
Given the remarkable success of diffusion models in various tasks, we adopt a diffusion-based approach for generating motions of both hands. Unlike PianoMotion10M~\cite{gan2025pianomotion10mdatasetbenchmarkhand}, which treats both hands as a whole and employs a single diffusion model for hand motion generation, our method decouples the generation process by utilizing two independent diffusion models to generate left and right hand gestures separately. 
Diffusion models are a class of generative models that progressively transform data into noise and then reverse this process to generate samples from a learned distribution.

In the forward process, a fixed Markov chain gradually adds Gaussian noise to the original data until it becomes nearly indistinguishable from pure noise. Mathematically, starting with a data sample $\mathbf{x}_0 \sim q(\mathbf{x}_0)$ , the forward process is defined as:
\begin{equation}
    \label{eq:diff_forard}
    q(\mathbf{x}_t | \mathbf{x}_{t-1}) = \mathcal{N}(\mathbf{x}_t; \sqrt{1-\beta_t}\, \mathbf{x}_{t-1}, \beta_t \mathbf{I}),
\end{equation}
where $\beta_t$ is a small constant controlling the noise level at each timestep. According to the property of Markov chains, $\mathbf{x}_t$ can be directly obtained using the following formula:
\begin{equation}
    \label{eq:x_t_0}
    \mathbf{x}_t = \sqrt{\bar{\alpha}_t} \mathbf{x}_0 + \epsilon \sqrt{1 - \bar{\alpha}_t},
\end{equation}
where $\alpha_t = 1-\beta_t$ and $\bar{\alpha}_t = \prod_{s=1}^{t} \alpha_s$.
The reverse process, which is learned using a neural network, is designed to gradually reconstruct the original data from a noisy sample by approximating the true reverse distribution. This process involves predicting and removing the added noise step by step,  reversing the forward diffusion process. By leveraging a parameterized neural network, the model learns to estimate the denoising transformation at each timestep, ultimately generating high-fidelity samples that closely resemble the original data distribution:
\begin{equation}
    \label{eq:diff_denoise}
    p_{\theta}(\mathbf{x}_{t-1} | \mathbf{x}_t, \mathbf{c}) = \mathcal{N}(\mathbf{x}_{t-1}; \mu_{\theta}(\mathbf{x}_t, t, \mathbf{c}), \Sigma_{\theta}(\mathbf{x}_t, t, \mathbf{c})),
\end{equation}
where the neural network predicts the mean $\mu_\theta(\cdot)$ and the variance $\Sigma_\theta(\cdot)$ based on $\mathbf{x}_t$, timestep $t$, and the context information $\mathbf{c}$. Specifically, $\mathbf{c}$ is constructed via the concatenation of audio feature and positional prediction information. 
To accomplish the reverse process of the diffusion model, we need to construct and optimize the neural network to predict unknown variables such as the initial value $\mathbf{x}_0$ and the noise $\epsilon$. Nichol et al.~\cite{dhariwal2021diffusionmodelsbeatgans, nichol2021improveddenoisingdiffusionprobabilistic} used this network to predict the noise $\epsilon$, while Ramesh et al.~\cite{ramesh2022hierarchicaltextconditionalimagegeneration, tevet2022humanmotiondiffusionmodel,chen2024enablingsynergisticfullbodycontrol} employed it to predict the initial value $\mathbf{x}_0$. Salimans et al.~\cite{salimans2022progressivedistillationfastsampling} introduced a novel parameterization method for diffusion models, incorporating a parameter $\mathbf{v}$, which is defined as:
\begin{equation}
    \label{eq:v}
    \mathbf{v_t}\equiv\sqrt{\bar\alpha_t}\epsilon-\sigma_t\mathbf{x_0},
\end{equation}
where $\sigma_t = \sqrt{1 - \bar{\alpha}_t}$ and $\epsilon \sim \mathcal{N}(\mathbf{0}, \mathbf{I})$. This approach has demonstrated superior performance in multiple studies. Therefore, we adopt this method here and define the loss function as:
\begin{equation}
    \label{eq:loss_mse}
    \mathcal{L}_{\text{MSE}} = \mathbb{E}_{t \in [1, T], \mathbf{x}_0 \sim q(\mathbf{x}_0), \epsilon \sim \mathcal{N}(\mathbf{0}, \mathbf{I})} \left[ \|\mathbf{v} - \mathbf{v}_\theta(\mathbf{x}_t, t, \mathbf{c})\|^2 \right].
\end{equation}

Considering the independence and asymmetry of left and right hand movement patterns in piano performance, we adopt a decoupled approach by employing two independent diffusion models to separately model the mapping between hand gestures and audio. During training, these two models are optimized in parallel. For a given time step $t$, we sample the original motion data $\mathbf{x}_0$, which consists of hand motion components $\mathbf{x}_{0L}$ and $\mathbf{x}_{0R}$. Each model then samples noise independently from a standard normal distribution and feeds it into its respective denoising model for training, learning to reconstruct realistic hand gestures gradually.
Our method uses a dual-stream diffusion model to generate left and right hand gestures during inference independently. Each stream samples initial noise from a standard normal distribution and refines it through iterative denoising to produce realistic gestures$\tilde{\mathbf{x}}_0$ for both hands. The coordinated gesture generation process is outlined in Algorithm~\ref{alg:piano_motion_left}, which details the inference steps for the left hand, while the right hand follows a symmetric approach.
\begin{figure}[htbp] 
  \includegraphics[width=0.47\textwidth]{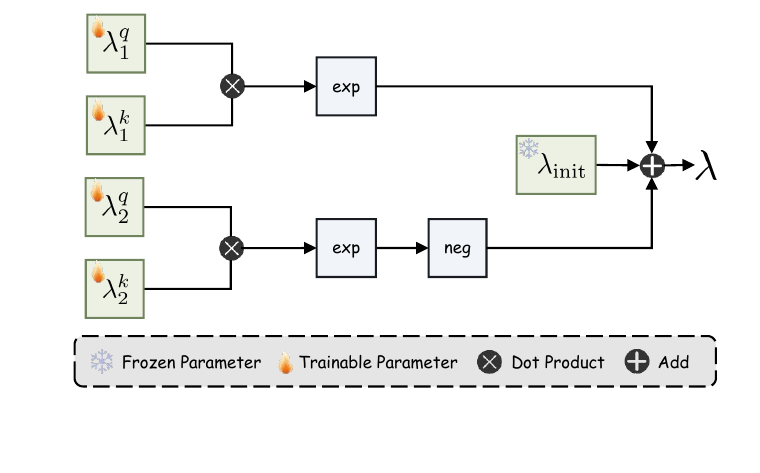}
  \caption{The way to compute the parameter $\lambda$.}
  \label{fig:lambda}
\end{figure}
\begin{algorithm}[htbp] 
\caption{Dual-Stream Coordinated Diffusion Inference}
\label{alg:piano_motion_left}
\begin{algorithmic}[1] 
    \Require Piano audio segment $A$
    \Ensure Generated left and right gesture sequences: $\Theta_L$, $\Theta_R$
    
    \Statex \textbf{Modules:}
    \Statex \quad $\Phi_a$: Audio Feature Extractor (e.g., HuBERT)
    \Statex \quad $\Phi_r^L, \Phi_r^R$: Audio Feature Refiner for each hand
    \Statex \quad $\Phi_{A2P}^{L}, \Phi_{A2P}^{R}$: Audio-to-Position Predictor for each hand
    \Statex \quad $\Phi_{A2M}^{L}, \Phi_{A2M}^{R}$: Audio-to-Motion Generator for each hand
    \Statex \quad $\text{HCAA}(\cdot, \cdot)$: Feature Interaction Module
    
    \Statex \textbf{Parameters:}
    \Statex \quad $T$: Total diffusion timesteps
    \Statex \quad $\alpha_t, \beta_t$: Diffusion schedule
    
    \Function{GenerateMotion}{$A$}
        \State $f_a \gets \Phi_a(A)$ 
        \State $(P_L, P_R) \gets (\Phi_{A2P}^{L}(f_a), \Phi_{A2P}^{R}(f_a))$ 
        \State $(\mathbf{x}_{T}^{L}, \mathbf{x}_{T}^{R}) \sim \mathcal{N}(0, \mathbf{I})$ 
        
        \Statex \hspace{\algorithmicindent} \parbox[t]{0.85\linewidth}{
        \textit{/* This loop illustrates the left-hand generation process; the right-hand has a symmetric implementation. */}}

        \For{$t = T$ \textbf{to} $1$}
            \State $\epsilon \sim \mathcal{N}(0, \mathbf{I})$ 
            \textbf{if} $t > 1$ \textbf{else} $\epsilon = 0$
            \State $\mathbf{c} \gets \text{concat}(\Phi_r^L(f_a), P_L, P_R)$ 
            \State $\mathbf{v}_{\theta}^{L} \gets \Phi_{A2M}^{L}(\mathbf{x}_{t}^{L}, t, \mathbf{c}, \text{HCAA}(\mathbf{x}_{t}^{L}, \mathbf{x}_{t}^{R}))$ 
            \State $\hat{\mathbf{x}}_{0}^{L} \gets \sqrt{\bar{\alpha}_t} \mathbf{x}_{t}^{L} - \sqrt{1-\bar{\alpha}_t} \mathbf{v}_{\theta}^{L}$
            \State $\text{mean}^L \gets \frac{\beta_t\sqrt{\bar{\alpha}_{t-1}}}{1-\bar{\alpha}_t} \hat{\mathbf{x}}_0^L + \frac{(1-\bar{\alpha}_{t-1})\sqrt{\alpha_t}}{1-\bar{\alpha}_t} \mathbf{x}_t^L$
            \State $\text{log\_var}^L \gets \log\left(\frac{(1-\bar{\alpha}_{t-1})\beta_t}{1-\bar{\alpha}_t}\right)$
            \State $\mathbf{x}_{t-1}^{L} \gets \text{mean}^L + e^{\frac{\text{log\_var}^L}{2}} \cdot \epsilon$
        \EndFor
        
        \State $(\Theta_L, \Theta_R) \gets (\mathbf{x}_{0}^{L}, \mathbf{x}_{0}^{R})$
        \State \Return $(\Theta_L, \Theta_R)$
    \EndFunction
\end{algorithmic}
\end{algorithm}

\subsection{Hand-Coordinated Asymmetric Attention}
While the left and right hands have their own independence, they still need to work closely together to maintain the fluency and coordination of the playing. Therefore, we not only use a decoupling strategy to generate separate motions for each hand but also introduce a feature interaction module. 
Inspired by Ye~et~al.~\cite{ye2024differential}, we introduce a Hand-Coordinated Asymmetric Attention (HCAA) mechanism to enhance information exchange between the left and right hand gesture generators. During inference, the diffusion model gradually removes noise through a denoising process.
However, at any given timestep during training or inference, the feature representations still contain certain noise interference, which causes the model to focus on irrelevant contextual information. 
Aiming to mitigate the impact of the aforementioned noise, we perform differential computation on the feature representations of the left and right hands in the intermediate layers of the diffusion model asymmetrically:
\begin{equation}
\label{eq:DAttn}
\operatorname{Attn}_s= \text{softmax}\left(\frac{Q_s K_s^T}{\sqrt{d_k}} \right)V_s \\ - \lambda_s\ \text{softmax}\left(\frac{Q_{\bar{s}} K_{\bar{s}}^T}{\sqrt{d_k}} \right)V_{\bar{s}},
\end{equation}
where $s \in \{\text{left}, \text{right}\}$, and $\bar{s}$ denotes the opposite hand of $s$. Here, $Q_{s}$, $K_{s}$, $V_{s}$ are the Query, Key, and Value matrices from the intermediate features of stream $s$, and $d_k$ denotes the dimensionality of the input features. $\lambda_{s}$ is the parameter $\lambda$ for stream $s$, which is a learnable coefficient controlling the degree of cross-hand suppression.
As shown in Fig.~\ref{fig:lambda}, we calculate \( \lambda \) as follows:
\begin{equation}
\lambda = \lambda_{1} - \lambda_{2} + \lambda_{\text{init}},
\end{equation}
where \( \lambda_{\text{init}} \) is a constant offset. This allows the model to automatically adapt during training and capture the inherent differences between the hands.
The values of \( \lambda_{1} \) and \( \lambda_{2} \) are computed as:
\begin{equation}
\begin{aligned}
\lambda_{1} &= \exp \left( \sum_{i} \left( \lambda_{1,i}^{q} \cdot \lambda_{1,i}^{k} \right) \right), \\
\lambda_{2} &= \exp \left( \sum_{i} \left( \lambda_{2,i}^{q} \cdot \lambda_{2,i}^k \right) \right),
\end{aligned}
\end{equation}
where \( \lambda_{1}^{q}, \lambda_{1}^{k}, \lambda_{2}^{q}, \) and \( \lambda_{2}^{k} \) are learnable parameters.
This computation is analogous to a differential amplifier in electrical engineering, which extracts differences between two input signals to suppress symmetric common-mode noise and highlight meaningful asymmetric signals. Since the noise distributions in the left and right hand diffusion models are similar, this asymmetric attention mechanism allows the model to filter out shared noise effectively. As a result, the learned features become more robust and expressive.

By incorporating this mechanism, the model not only better captures the interactions between the left and right hands but also improves the stability and consistency of the generated motions. Experimental results demonstrate that, compared to conventional attention mechanisms, asymmetric attention achieves superior performance across multiple evaluation metrics and significantly enhances the quality of hand motion generation, resulting in more coordinated and natural synthesized piano-playing gestures.

\definecolor{myblue}{RGB}{230, 230, 248}

\begin{table*}[htbp]
\caption{Quantitative evaluation of our decoupled hand motion generation method and existing models on PianoMotion10M.}
\label{tab:quan}
\renewcommand\arraystretch{1.2}
\resizebox{1\linewidth}{!}{
\begin{tabular}{c|c|cccc|cccc|c}
\Xhline{1pt}
\multirow{2}{*}{\textbf{Method}} & \multirow{2}{*}{\textbf{Venue}} & \multicolumn{4}{c|}{\textbf{Right Hand}}& \multicolumn{4}{c|}{\textbf{Left Hand}}& \multirow{2}{*}{\textbf{FID$\downarrow$}}\\ \cline{3-6} \cline{7-10}
\multicolumn{1}{c|}{} &  & \textbf{FGD$\downarrow$} & \textbf{WGD$\downarrow$} & \textbf{PD$\downarrow$} & \textbf{Smooth$\downarrow$} & \textbf{FGD$\downarrow$} & \textbf{WGD$\downarrow$} & \textbf{PD$\downarrow$} & \textbf{Smooth$\downarrow$} & \\ \hline
EmoTalk~\cite{peng2023emotalk} & ICCV 2023 & 0.360 & 0.259 & 0.033 & 0.313 & 0.445 & 0.232 & 0.044 & 0.353 & 4.645 \\
\multicolumn{1}{c|}{LivelySpeaker~\cite{zhi2023livelyspeaker}} & ICCV 2023 & 0.535 & 0.249 & 0.030 & 0.334 & 0.538 & 0.220 & 0.038 & 0.406 & 4.157 \\ 
PianoMotion~\cite{gan2025pianomotion10mdatasetbenchmarkhand} & ICLR 2025 & 0.351 & 0.244 & 0.030 & 0.205 & 0.372 & 0.217 & 0.037 & 0.248  & 3.281 \\ \hline
\ModelName~(ours) & MM 2025 & \cellcolor{myblue}0.293 & \cellcolor{myblue}0.242 & \cellcolor{myblue}0.028 & \cellcolor{myblue}0.198 & \cellcolor{myblue}0.309 & \cellcolor{myblue}0.215 & \cellcolor{myblue}0.035 & \cellcolor{myblue}0.229 & \cellcolor{myblue}3.012
\\ \Xhline{1pt} 
\end{tabular}%
}
\end{table*}

\begin{figure*}[t] 
  \includegraphics[width=0.99\textwidth]{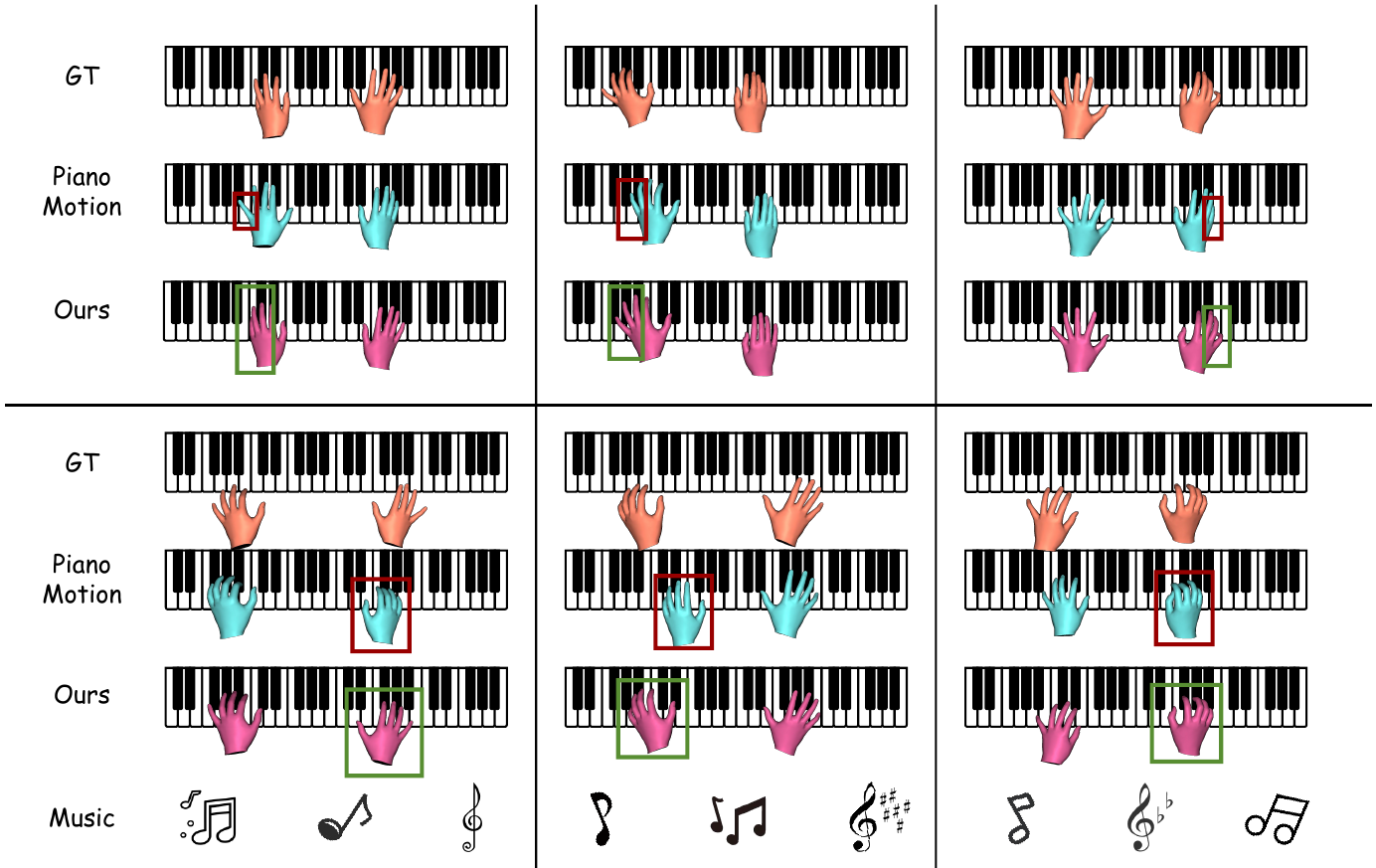}
  \caption{Illustration of the qualitative results. We visualize the generated gestures across frames using different methods, where GT represents Ground Truth and PianoMotion denotes PianoMotion10M. The anomalous generated gestures in the figure are highlighted with red boxes. In contrast, the relatively natural generated gestures are highlighted with green boxes.
  }
  \label{fig:quality}
\end{figure*}
\section{Experiments}
\subsection{Experimental Settings}
\textbf{Dataset.} We use the PianoMotion10M~\cite{gan2025pianomotion10mdatasetbenchmarkhand} dataset, the currently largest large-scale piano motion dataset. PianoMotion10M consists of 1,966 piano performance videos with a total duration of 116 hours, along with synchronized MIDI files and 10 million corresponding hand pose annotations. These videos are segmented into 16,739 independent 30-second clips, with 7,519 used for training, 821 for validation, and 8,399 for testing. The dataset includes recordings from 14 different subjects, capturing diverse playing styles and ensuring broad applicability.

\begin{table*}[htbp]
\caption{
Ablation study on the proposed modules. Here, DN represents Decoupled Noise, indicating whether the noise used in the two diffusion models is disentangled. PS stands for Position Sharing, and FI denotes Feature Interaction. Concat refers to Concatenation, CA represents Cross Attention, and HCAA stands for Hand-Coordinated Asymmetric Attention.}
\label{tab:benchmark}
\renewcommand\arraystretch{1.2}
\resizebox{1\linewidth}{!}{
\begin{tabular}{c|c|c|c|cccc|cccc|c}
\Xhline{1pt}
\multirow{2}{*}{\textbf{Method}}& \multirow{2}{*}{\textbf{DN}} & \multirow{2}{*}{\textbf{PS}} & \multirow{2}{*}{\textbf{FI}} & \multicolumn{4}{c|}{\textbf{Right Hand}}& \multicolumn{4}{c|}{\textbf{Left Hand}}& \multirow{2}{*}{\textbf{FID$\downarrow$}}\\ \cline{5-8} \cline{9-12}
 & & &  & \textbf{FGD$\downarrow$} & \textbf{WGD$\downarrow$} & \textbf{PD$\downarrow$} & \textbf{Smooth$\downarrow$} & \textbf{FGD$\downarrow$} & \textbf{WGD$\downarrow$} & \textbf{PD$\downarrow$} & \textbf{Smooth$\downarrow$} & \\ \hline

ours &  & \checkmark & HCAA & 0.703 & \cellcolor{myblue}0.222 & 0.028 & 0.247 & 0.583 & 0.216 & 0.035 & 0.299 & 9.267 \\ \hline

ours & \checkmark & & - & 0.368 & 0.240 & 0.028 & 0.200 & 0.362 & 0.215 & 0.035 & 0.237 & 3.232 \\

ours & \checkmark & \checkmark & - & 0.335 & 0.244 & 0.028 & 0.200 & 0.336 & 0.215 & 0.035 & 0.233 & 3.193 \\ \hline

ours & \checkmark  & \checkmark & Concat & 0.315 & 0.241 & 0.028 & \cellcolor{myblue}0.196 & 0.337 & 0.217 & 0.035 & 0.242 & 3.197 \\

ours & \checkmark & \checkmark & CA & 0.311 & 0.241 & 0.028 & 0.200 & 0.326 & 0.217 & 0.035 & 0.241 & 3.055 \\

ours & \checkmark & \checkmark & HCAA & \cellcolor{myblue}0.293 & 0.242 & \cellcolor{myblue}0.028 & 0.198 & \cellcolor{myblue}0.309 & \cellcolor{myblue}0.215 & \cellcolor{myblue}0.035 & \cellcolor{myblue}0.229 & \cellcolor{myblue}3.012 
\\ \Xhline{1pt} 
\end{tabular}%
}
\end{table*}

\textbf{Metrics.} To comprehensively evaluate the performance of the proposed method, we follow the evaluation metrics used in PianoMotion10M to assess the realism and effectiveness of the generated hand motions from piano audio input. These evaluation metrics help evaluate the quality of the generated motions and also reveal how the gestures correspond to the musical structure of the piano performance.

\begin{itemize}[leftmargin=*,noitemsep,nolistsep]
    \item \textbf{Frechet Inception Distance (FID)}: FID measures the similarity between predicted gestures and real data in the latent space by computing the Frechet distance between their latent feature vectors. This metric evaluates the distributional difference between generated results and real data.

    \item \textbf{Frechet Gesture Distance (FGD)}: Unlike FID, FGD is used to measure the difference between the predicted gestures and ground truth  in the MANO\cite{romero2017embodied} parameter space for a single hand. This metric helps evaluate the similarity of single-hand gestures.

    \item \textbf{Wasserstein Gesture Distance (WGD)}: WGD~\cite{Rubner2000TheEM} quantifies the difference between two gesture distributions by computing the minimum “cost” required for optimal transport between probability distributions. Each distribution is modeled using a Gaussian Mixture Model (GMM)~\cite{kolouri2017slicedwassersteindistancelearning} to accurately capture the complex characteristics of gestures.
    
    \item \textbf{Position Distance (PD)}: PD measures the discrepancy between predicted and ground-truth hand positions using the L2 (Euclidean) distance. This metric directly reflects the spatial accuracy of the generated hand trajectories, indicating how closely the predicted positions align with those of real performance gestures.
    
    \item \textbf{Smoothness}: Smoothness is measured by calculating the acceleration of each joint. However, static hand positions exhibit the highest smoothness, which contradicts the desired outcome. To address this, we use the ground truth acceleration as a reference and employ relative acceleration as the smoothness metric. 
\end{itemize}

\textbf{Implementation Details.} We select HuBERT~\cite{hsu2021hubertselfsupervisedspeechrepresentation} as the audio feature extractor and utilize multiple transformer layers~\cite{vaswani2017attention} as the audio refiner. 
Specifically, the HuBERT feature extractor produces features with a dimensionality of 1024, while the audio refiner adheres to a standard Transformer architecture, with both the encoder and decoder comprising four layers each.
Each training sample has a duration of 8 seconds with a frame rate of 30 frames per second, yielding a sequence of 240 frames. The feature dimension for both the left and right hand joints is 48.
For the diffusion model, we use time steps of $T=1000$ and a linearly increasing variance schedule of $\beta_1=1e-4$ to $\beta_T = 0.02$. Our denoising model adopts a 4-layer U-Net architecture, with each layer having dimensions of 256, 512, 1024, and 2048, respectively. We use an Adam optimizer with a learning rate of $2.5e-5$. All experiments were conducted on a single NVIDIA RTX 4090 GPU, with a training duration of approximately 4 days.

\subsection{Experimental Results}
\textbf{Quantitative Results.} 
Tab.~\ref{tab:quan} presents a quantitative comparison between the baseline method and our proposed approach across multiple evaluation metrics. During the evaluation, we strictly adhered to the same dataset split and hyperparameter settings as PianoMotion10M~\cite{gan2025pianomotion10mdatasetbenchmarkhand} to ensure a fair comparison. The results for EmoTalk~\cite{peng2023emotalk} and LivelySpeaker~\cite{zhi2023livelyspeaker} were obtained by Gan et al.~\cite{gan2025pianomotion10mdatasetbenchmarkhand} through the application of their method to the corresponding tasks.
Specifically, EmoTalk~\cite{peng2023emotalk} generates poses directly from audio, while LivelySpeaker~\cite{zhi2023livelyspeaker} utilizes a multi-layer perceptron (MLP)-based diffusion framework for motion generation. The results indicate that our method produces a more realistic and coherent generation of hand gestures, surpassing the baseline in key aspects.
Compared to the baseline model, our method achieves a significant reduction in the two primary metrics: FGD and FID. This indicates a more accurate modeling of the correspondence between audio and hand motions, as well as more realistic and natural gesture generation. Additionally, consistent improvements in other metrics (e.g., WGD, PD, and Smooth) further demonstrate that our approach excels not only in overall motion quality but also in spatial precision and smooth temporal transitions.

\textbf{Qualitative Results.} 
We compare the generated motions from our method with those from PianoMotion10M~\cite{gan2025pianomotion10mdatasetbenchmarkhand}. 
As shown in Fig.~\ref{fig:quality}, our approach is capable of producing piano-playing gestures that closely resemble real hand movements, whereas the PianoMotion10M method exhibits issues in hand positioning, overall motion consistency, and fine details. 
For instance, in the top three frames of Fig.~\ref{fig:quality}, the gestures generated by PianoMotion10M display significant errors in finger details. In the bottom three frames, the hand poses produced by PianoMotion10M demonstrate inaccuracies in either the trajectory of hand movements or their positioning. These findings suggest that our decoupled strategy for position prediction and motion generation enables the model to concentrate more effectively on the location and intricate details of each hand. Furthermore, the introduced feature interaction mechanism improves coordination between both hands, resulting in more natural and synchronized gesture generation.

\subsection{Ablation Studies}
In this section, we evaluate each design in our framework to verify its effectiveness. Tab.~\ref{tab:ablation} presents the results of our ablation studies on different proposed components, including the effect of decoupled noise, the feature interaction module, the position sharing strategy, and our proposed Hand-Coordinated Asymmetric Attention.

\textbf{Effect of decoupled noise.} 
During both training and inference, using the same noise input for the left and right hand generation models results in suboptimal performance. Although the two hands must cooperate during the piano performance, they inherently exhibit distinct and independent motion patterns. Sharing identical noise across both hands imposes unnecessary constraints on the generation process, limiting the expressive independence of each hand and ultimately leading to unnatural or poorly coordinated movements. To address this issue, we introduce independent noise into the diffusion models for each hand. This design allows the model to better capture the unique motion dynamics of each hand, thereby enhancing the realism, diversity, and overall quality of the generated gestures.

\textbf{Effect of feature interaction module.} 
Incorporating the feature interaction module (HCAA) significantly enhances the performance of the motion generation model. This improvement arises from the fact that, although the movements of the left and right hands in piano performance exhibit a certain degree of independence, they are inherently interdependent in terms of rhythm, dynamics, and musical structure. Ignoring this coordination during generation may result in disjointed or unnatural gestures. The feature interaction module explicitly models the information exchange between the two hands, enabling the model to better capture the collaborative patterns inherent in piano playing. Consequently, it generates more coherent, expressive, and natural hand motion sequences.

\textbf{Effect of position sharing strategy.} 
At both the model training and inference stages for gesture generation, position prediction serves as a crucial input condition that significantly impacts the quality of the generated motions. In contrast to the approach where each hand's generation model only receives its own predicted positions, the position-sharing strategy, where both models utilize the predicted positions of both hands as input, yields noticeably better performance. This improvement can be attributed to the fact that, in piano performance, the spatial positions of the two hands are often closely coordinated. By sharing positional information between the two models, the system can more effectively capture the inter-hand spatial dependencies, resulting in the generation of more natural, synchronized, and coherent hand motions.

\textbf{Effect of HCAA.} 
Compared to simple feature concatenation and conventional cross-attention mechanisms, our proposed Hand-Coordinated Asymmetric Attention (HCAA) demonstrates superior performance in motion generation. This enhancement is because the intermediate features in the denoising process of diffusion models still retain a significant amount of noise. Traditional fusion methods, such as concatenation and cross-attention, are susceptible to introducing redundant or misleading information, which can compromise the quality of the generated motions. In contrast, our asymmetric attention mechanism calculates the difference between the attention responses of the two hands, effectively suppressing symmetric common-mode noise shared across both sides. This approach improves the robustness of the feature interaction process, allowing the model to better concentrate on the essential coordinated patterns between the hands, ultimately resulting in more stable, natural, and expressive motion synthesis.

\section{Limitations and Future Work}
Although our proposed dual-stream diffusion framework demonstrates strong performance in generating high-quality and coordinated piano hand motions, several challenges and opportunities for improvement remain. 

First, the current model operates solely based on audio input. However, in real-world piano performances, a performer's hand gestures are influenced by various factors beyond the auditory signal, such as personal playing style, sizes of hands and physical expression habits. To address this limitation, future work could explore incorporating multi-modal inputs, such as visual recordings, identity or style embeddings, and body posture cues. By fusing these additional modalities, the model could produce more personalized, expressive, and context-aware motions.

Second, the availability of large-scale paired piano-audio and 3D hand motion datasets is still limited. The scarcity of diverse training data constrains the model's ability to generalize to various playing styles or complex pieces. Moreover, existing datasets often lack detailed physical constraints, such as finger-key contact labels or joint-angle limits, which are essential for ensuring physically plausible and musically accurate motions. Future research may consider constructing or expanding datasets that include a broader range of musical genres, playing techniques, individual performers and explicit physical annotations. This would enable the model to learn a more nuanced mapping between audio and gesture dynamics.

Finally, although diffusion models are known for their superior generation quality, their inference process is inherently slow due to the need for iterative denoising steps. This limits their practicality in real-time or interactive applications. To mitigate this issue, future work could adopt accelerated sampling techniques within diffusion models—such as the VAE-based latent compression strategy used by Chen et al.~\cite{chen2023executingcommandsmotiondiffusion}, or even replace the diffusion architecture altogether with more efficient generative paradigms like Latent Consistency Models (LCMs)~\cite{luo2023latentconsistencymodelssynthesizing}. These alternatives hold the potential to significantly speed up the generation process without sacrificing output quality.

\section{Conclusion}
In this paper, we present a novel framework that revisits the conventional two-hand joint modeling paradigm commonly employed in piano motion generation. Our approach explicitly addresses the often-overlooked functional differentiation and asynchrony between the left and right hands by introducing a decoupled dual-stream synthesis strategy. To effectively capture both independent hand dynamics and their inter-hand coordination, we further propose the Hand-Coordinated Asymmetric Attention (HCAA) mechanism. This design facilitates more expressive, temporally consistent, and musically aligned motion generation while preserving overall hand coordination.
Extensive experiments conducted on the PianoMotion10M dataset~\cite{gan2025pianomotion10mdatasetbenchmarkhand} demonstrate that our method consistently outperforms existing state-of-the-art approaches across multiple evaluation metrics, including naturalness, temporal smoothness, spatial accuracy, and coordination. Ablation studies further validate the contributions of key components, namely independent noise injection, the position information sharing mechanism, and the HCAA module, to the overall quality of generation. This work provides a new perspective on audio-driven piano-hand motion synthesis and establishes a solid foundation for future research in fine-grained musical motion generation.

\section*{Acknowledgement}
This work was supported by the STI 2030-Major Projects under Grant 2021ZD0201404.

{
    \small
    \bibliographystyle{ieeenat_fullname}
    \bibliography{main}
}

\section{More Details}
\subsection{User Study}
\begin{figure}[ht] 
  \includegraphics[width=0.47\textwidth]{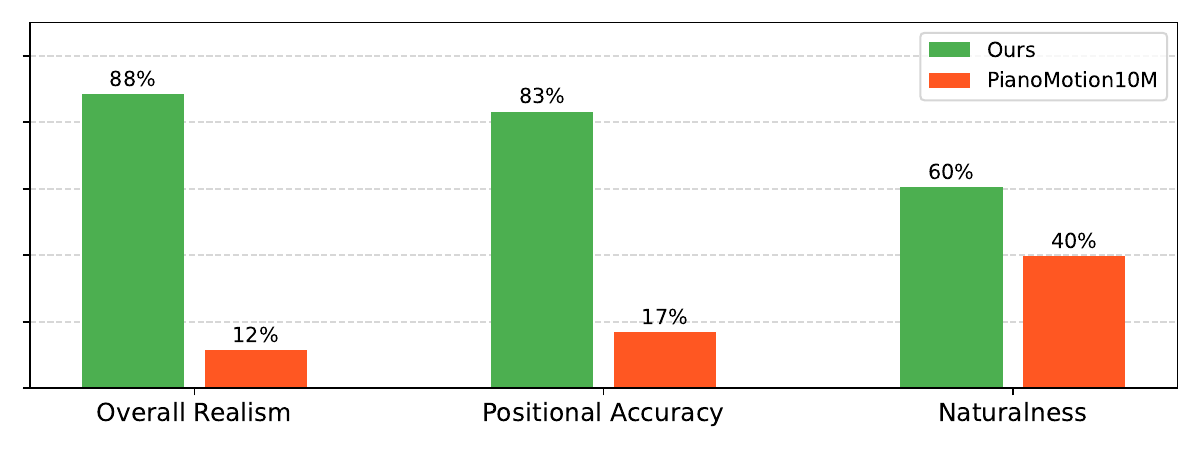}
  \caption{User study for synthesized motions. The current state-of-the-art and our methods are compared based on three evaluating factors. Each bar indicates the preference rate of our method over pianoMotion10M.}
  \label{fig:user_study}
\end{figure}

As shown in Fig.~\ref{fig:user_study}, we conducted a user study involving 15 participants to evaluate the quality of the generated hand motions. The evaluation was based on the following three dimensions:
\begin{enumerate}
    \item \textbf{Overall realism}: Participants were asked to assess how closely the generated motions resemble real hand movements at a global level.
    \item \textbf{Positional accuracy}: This criterion considers whether the positions of the generated hands align well with those in real performances, particularly whether the fingers press the correct keys.
    \item \textbf{Naturalness}: Participants evaluated whether the generated motions exhibit structural plausibility, including any unnatural joint deformations or unrealistic finger-to-key interactions, such as awkward twisting or stiff contact gestures.
\end{enumerate}

From the results, our method received more favorable ratings compared to PianoMotion10M, indicating a clear preference for the motions generated by our model across all three dimensions.

\subsection{More Experiment of $\lambda$}

\definecolor{myblue}{RGB}{230, 230, 248}

\begin{table*}[htbp]
\caption{Ablation study on $\lambda_{\text{init}}$. We conducted experiments with different initial values of the parameter $\lambda_{\text{init}}$ to evaluate its impact on training.}
\label{tab:lambda}
\renewcommand\arraystretch{1.2}
\resizebox{1\linewidth}{!}{
\begin{tabular}{c|c|cccc|cccc|c}
\Xhline{1pt}
\multirow{2}{*}{\textbf{Method}} & \multirow{2}{*}{$\lambda_{\text{init}}$} & \multicolumn{4}{c|}{\textbf{Right Hand}}& \multicolumn{4}{c|}{\textbf{Left Hand}}& \multirow{2}{*}{\textbf{FID$\downarrow$}}\\ \cline{3-6} \cline{7-10}
 &  & \textbf{FGD$\downarrow$} & \textbf{WGD$\downarrow$} & \textbf{PD$\downarrow$} & \textbf{Smooth$\downarrow$} & \textbf{FGD$\downarrow$} & \textbf{WGD$\downarrow$} & \textbf{PD$\downarrow$} & \textbf{Smooth$\downarrow$} & \\ \hline

ours & 0.1 & 0.300 & 0.242 & 0.029 & 0.200 & 0.342 & 0.213 & 0.034 & 0.242 & 3.138 \\

ours & 0.3 & 0.303 & 0.241 & 0.029 & 0.200 & 0.332 & 0.213 & 0.034 & 0.235 & 3.079 \\

ours & 0.5 & 0.306 & 0.242 & 0.029 & 0.197 & 0.331 & 0.213 & 0.034 & 0.239 & 3.160

\\ \Xhline{1pt} 
\end{tabular}%
}
\end{table*}

In this section, we conduct additional experiments to investigate the impact of the $\lambda$ parameter, specifically focusing on the initialization value $\lambda_{\text{init}}$, which is set as a fixed constant during training. As shown in Tab.~\ref{tab:lambda}, we test multiple values (e.g., 0.1, 0.3, 0.5) and observe that the model remains robust across different settings, with performance variations within $4\%$ across all evaluation metrics. These results indicate that our model is relatively robust to the choice of $\lambda_{\text{init}}$, and its performance is not overly sensitive to this hyperparameter. Based on empirical observations, we empirically select $\lambda_{\text{init}} = 0.78$, which consistently yields stable and strong performance in practice.

\subsection{More visualization cases}

\begin{figure*}[t] 
  \includegraphics[width=\textwidth]{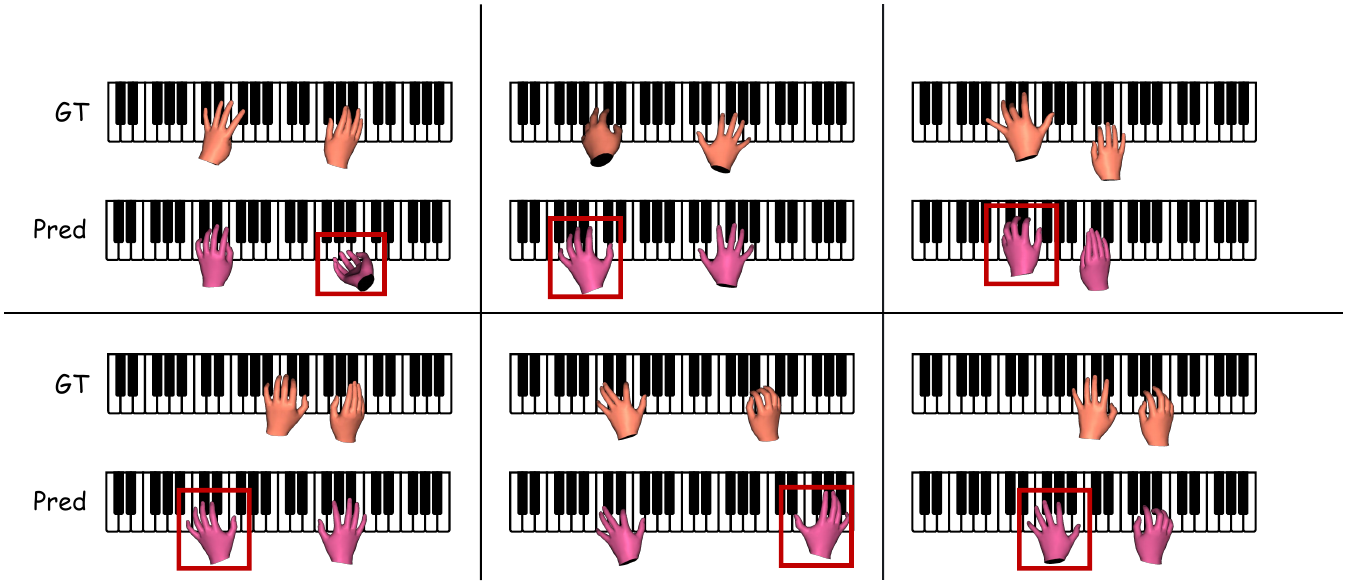}
  \caption{Visualizations of failure cases. We present several examples with abnormal results selected from the generated outputs to illustrate typical failure modes of our model. The anomalous generated gestures in the figure are highlighted with red boxes.}
  \label{fig:fail}
\end{figure*}

\begin{figure*}[t] 
  \includegraphics[width=\textwidth]{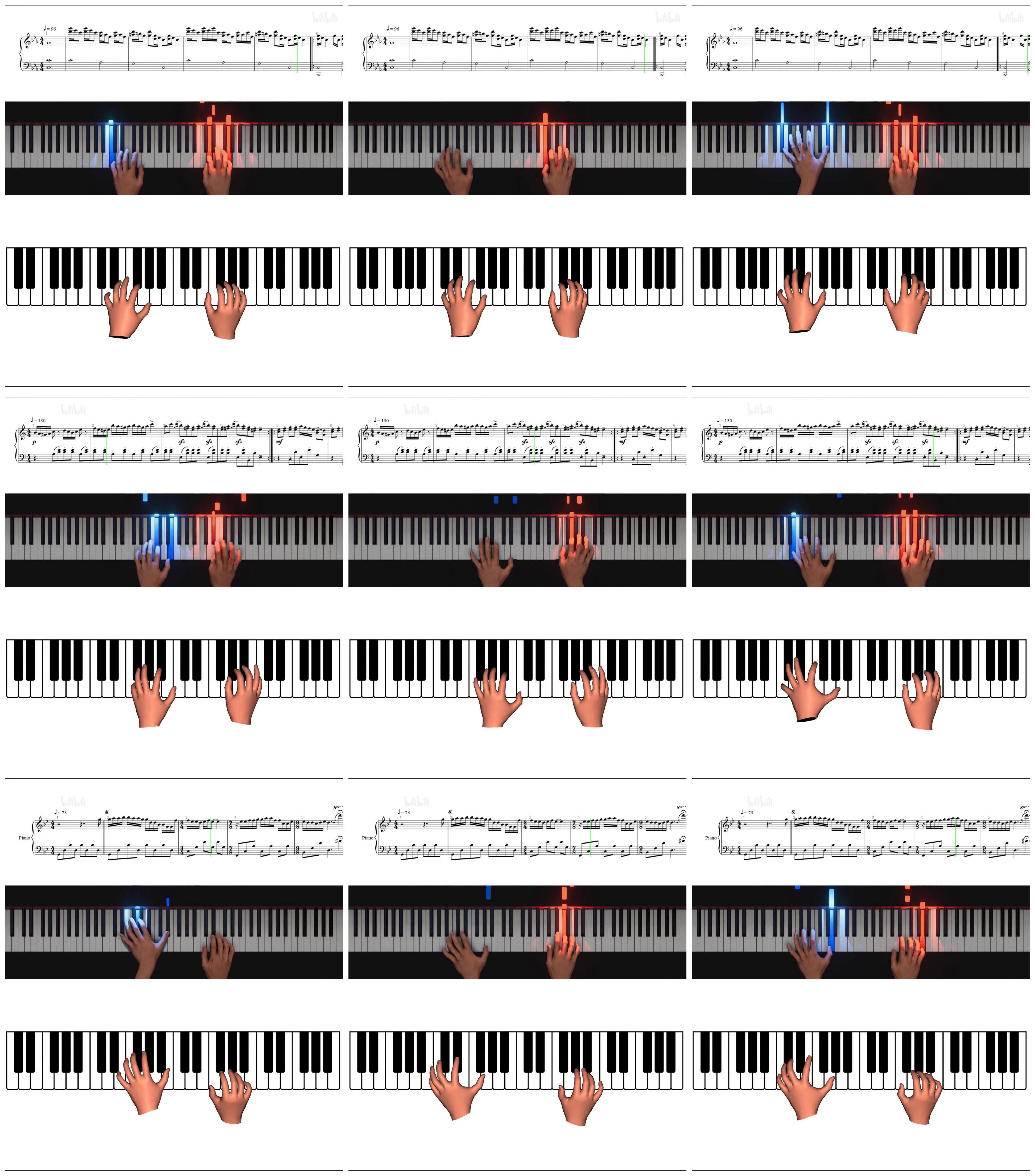}
  \caption{Test on wild data. We collected piano performance videos from social media — none of which appear in the training set (i.e., wild data with diverse genres and playing styles). We fed the corresponding audio into our model to generate hand motions.}
  \label{fig:wild}
\end{figure*}

In this section, we present several failure cases as well as visualizations on out-of-distribution data to highlight the limitations of our generative model and to demonstrate its generalization capability.

\textbf{Failure cases.}
Despite the strong performance of our method in the majority of cases, we observe that it can still produce abnormal results in certain challenging scenarios—particularly when the piano audio exhibits rapid tempo changes. As illustrated in Fig.~\ref{fig:fail}, the top three frames show abnormalities in hand posture, while the bottom three frames display positional inaccuracies. These issues highlight the inherent complexity of mapping audio signals to fine-grained hand motions. Such cases suggest that there remains significant room for improvement in modeling nuanced hand dynamics, especially under complex musical conditions, and we consider this an important direction for future work.

\textbf{Wild data.}
To evaluate the generalization ability of our method, we collected a set of piano performance videos with diverse styles from social media. These videos were not seen during training. We extracted the corresponding audio and fed it into our model to generate hand motion sequences. As shown in Fig.~\ref{fig:wild}, the generated motions closely resemble the ground truth, indicating that our method generalizes well to unseen, in-the-wild data with varying genres and playing styles.

In addition, we conducted a user study to assess the quality of these generated results. Participants rated the motions on a 10-point scale across three dimensions. The model received average scores of 6.9 for overall realism, 7.0 for positional accuracy, and 7.7 for naturalness, further demonstrating the effectiveness and generalizability of our approach.

\end{document}